\documentclass[preprint,3p,onecolumn,numbers]{elsarticle}

\usepackage{amsthm}
\usepackage{amsmath}
\usepackage{verbatim}
\usepackage{hyperref}
\usepackage{float}
\usepackage{bm,bbm}
\usepackage{subcaption}
\usepackage{url}
\usepackage{pifont}
\usepackage{amssymb} 
\usepackage{xcolor}

\journal{Journal of Economic Dynamics and Control}
\date{}

\begin{document}
\begin{frontmatter}

\title{A calibrated model of debt recycling with interest costs and tax shields: viability under different fiscal regimes and jurisdictions}

\author[1]{Carlo von der Osten}
\author[1]{Sabrina Aufiero}
\author[2]{Pierpaolo Vivo}
\author[1,3]{Fabio Caccioli}
\author[1,4]{Silvia Bartolucci}

\affiliation[1]{organization={Department of Computer Science, University College London},
            addressline={66-72 Gower Street}, 
            city={London},
            postcode={WC1E 6BT}, 
            country={UK}}
\affiliation[2]{organization={Department of Mathematics, King's College London},
            addressline={Strand}, 
            city={London},
            postcode={WC2R 2LS}, 
            country={UK}}
\affiliation[3]{organization={Systemic Risk Centre, London School of Economics and Political Sciences},
            city={London},
            postcode={WC2A 2AE}, 
            country={UK}}
\affiliation[4]{Corresponding author: s.bartolucci@ucl.ac.uk}

\begin{abstract}
    Debt recycling is a leveraged equity management strategy in which homeowners use accumulated home equity to finance investments, applying the resulting returns to accelerate mortgage repayment. We propose a novel framework to model equity and mortgage dynamics in presence of mortgage interest rates, borrowing costs on equity-backed credit lines, and tax shields arising from interest deductibility. The model is calibrated on three jurisdictions — Australia, Germany, and Switzerland — representing diverse interest rate environments and fiscal regimes. Results demonstrate that introducing positive interest rates without tax shields contracts success regions and lengthens repayment times, while tax shields partially reverse these effects by reducing effective borrowing costs and adding equity boosts from mortgage interest deductibility. Country-specific outcomes vary systematically, and rental properties consistently outperform owner-occupied housing due to mortgage interest deductibility provisions.
\end{abstract}

\end{frontmatter}

\section{Introduction}\label{sec: intro}

Debt recycling represents one of the most sophisticated yet risky strategies available to homeowners seeking to accelerate mortgage repayment. The strategy leverages accumulated home equity to finance investments in risky assets, effectively converting non-deductible mortgage debt into tax-efficient investment debt \cite{aufiero2025phase,ATO2025a}. While equity progressively builds up as the mortgage is repaid monthly, mortgage holders may obtain another loan backed by this equity to invest in income-producing assets. The wealth generated by successful investments can then be channelled toward faster mortgage repayment. However, the strategy carries substantial risk: fluctuations in both the housing market and investment performance may lead to rapid default, as both the mortgage and the liquidity loan are secured against the same property.

Aufiero et al. (2025) \cite{aufiero2025phase} provide a parsimonious dynamical model of debt recycling for the time evolution of equity and mortgage balance as functions of the loan-to-value (LTV) ratio, the performance of the house market, and the returns of the risky investment. Their model reveals a rich behavioural landscape characterised by distinct scenarios: \textit{strongly successful} outcomes where the mortgage is repaid faster than under standard monthly repayment strategies, \textit{weakly successful} phases where repayment is slower than via traditional approaches, \textit{default} outcomes when equity vanishes before the mortgage is fully repaid, and \textit{permanent remortgaging} phases where continuous refinancing prevents ultimate mortgage repayment. The model demonstrates high sensitivity to initial mortgage-to-equity ratios and scheduled repayment amounts. These findings establish clear phase boundaries that delineate when debt recycling may be recommended or discouraged based on the borrower's circumstances and market conditions.

However, the base model abstracts away from real-world features of the process that may alter the cash flows driving the equity and mortgage dynamics in every period. First, households face interest charges on both their primary mortgage and any equity-backed borrowing used for investment purposes. These interest costs directly reduce the available equity, while increasing the burden of debt service. Furthermore, in many tax jurisdictions deductions are provided for the interest paid on investment-related borrowing, effectively creating a tax shield that partially offsets borrowing costs. In some countries, mortgage interest on income-producing properties may also be tax-deductible, while interest on primary residences typically is not. The omission of these interest rate effects and tax considerations from the base model represents a significant gap between theory and practice, as these factors materially affect both the feasibility and attractiveness of debt recycling strategies in real-world applications.

This study builds on the base debt-recycling model to incorporate mortgage interest rates, borrowing costs on equity-backed loans, and the associated tax shields that arise under different fiscal regimes. By integrating these features into the original analytical framework, we provide a more realistic representation of the financial dynamics that households face when considering debt recycling strategies. The model's improvements preserve the tractability of the average process analysis, while introducing the actual costs and benefits that determine whether debt recycling accelerates or impedes mortgage repayment in practice.

Incorporating these features leads to a more detailed and accurate estimation of phase boundaries and outcome timing, refining their relevance for practical applications.

Country-specific calibration demonstrates that debt recycling viability is critically dependent on local financial and fiscal conditions. Strategies appearing prudent in one jurisdiction may prove hazardous in another, even for identical households. This jurisdictional heterogeneity has important implications for both household financial planning and cross-border regulatory policy. By calibrating our realistic debt recycling model to Australia, Germany, and Switzerland, we capture a representative range of institutional configurations: high rates with selective deductibility (Australia), moderate rates with limited deductibility (Germany), and low rates with broad deductibility (Switzerland). This approach allows us to quantify how institutional variation maps into outcome differences while holding household characteristics constant.

The remainder of this paper proceeds as follows. Sec. \ref{sec:lit review} reviews related works. Sec. \ref{sec: methods} reviews the base model by Aufiero et al. (2025) \cite{aufiero2025phase} and presents our analytical model incorporating interest rates and tax shields. It also describes the country-specific calibrations for Australia, Germany, and Switzerland, detailing the institutional parameters and data sources used. Sec. \ref{sec: Results} presents the main results. Sec. \ref{sec: concl} discusses the findings and their potential policy implications.

\section{Related Works}\label{sec:lit review}

Aufiero et al. (2025) \cite{aufiero2025phase} develop a discrete-time dynamical model of debt recycling where homeowners leverage accumulated equity to finance risky investments, while maintaining regular mortgage payments. The model tracks two coupled state variables, equity $E_t$ (the value of the portion of the house owned) and mortgage balance $M_t$ (remaining debt), governed by stochastic processes that incorporate investment returns, housing market fluctuations, and scheduled repayments. The authors identify four distinct outcome phases: \textit{strong success} (mortgage repaid faster than traditional amortisation),\textit{ weak success }(mortgage repaid but slower than traditional methods), \textit{default} (equity depleted before mortgage satisfaction), and \textit{permanent re-mortgaging} (neither boundary reached within finite horizons). Through eigenvalue analysis of the average transition matrix, they derive closed-form expressions for mean processes and construct phase diagrams over the $(p,s)$ parameter space, where $p$ denotes investment success probability and $s$ represents housing market drift. The base model demonstrates high sensitivity to the loan-to-value (LTV) ratio, investment risk factor, and initial mortgage-to-equity ratio. Sharp phase boundaries separate success from default regions, with small parameter changes that generate discontinuous jumps in outcomes, akin to first-order phase transitions in stochastic systems \cite{redner2001guide}. While the model provides elegant analytical results, it abstracts from interest costs and tax deduction considerations, treating debt as costless beyond scheduled principal repayments. 

A substantial empirical literature documents that interest rates materially affect mortgage default probabilities and household debt dynamics. Campbell and Cocco (2015) \cite{campbell2015model} develop a life-cycle model incorporating labor income, house price, inflation, and interest rate risk, demonstrating that mortgage premia and default decisions depend critically on whether borrowers face adjustable or fixed rates. Higher rates increase monthly payment burdens, tightening liquidity constraints and raising default thresholds even for borrowers with positive home equity.
Fuster and Willen (2017) \cite{fuster2017payment} provide empirical evidence that payment size, which scales directly with interest rates, substantially affects default propensities. Using loan modification programs that reduced required payments, they find that halving monthly payments lowers delinquency hazards by approximately $55\%$, even for deeply underwater borrowers: cash flow constraints drive default decisions. 

Tax treatment of mortgage and investment interest creates incentive structures that vary substantially across jurisdictions and property types. Poterba (1984) \cite{poterba1984tax} and Poterba and Sinai (2008) \cite{poterba2008tax} analyse the U.S. mortgage interest deduction, demonstrating that tax subsidies capitalise into house prices and alter optimal leverage choices. The deductibility of interest reduces the effective user cost of housing capital, encouraging higher leverage among itemizing taxpayers.
Empirical evidence on tax effects comes primarily from cross-sectional variation in tax rates or policy reforms. Dunsky and Follain (2000) \cite{dunsky2000tax} exploit U.S. state-level differences in marginal tax rates to estimate elasticities of mortgage demand with respect to after-tax interest costs, finding significant responses. Jappelli and Pistaferri (2007) \cite{jappelli2007people} use Italian tax reforms to identify causal effects of mortgage interest deductibility on homeownership rates and leverage choices.

The international comparison literature highlights substantial heterogeneity in tax treatment. European countries generally provide limited or no deductibility for owner-occupied mortgage interest, while investment property interest receives more favorable treatment \cite{drudi2009housing}. Australia represents an intermediate case where investment-related interest is broadly deductible but primary residence mortgage interest is not (Yates, 2011) \cite{yates2011housing}. Switzerland's approach of taxing imputed rental income while allowing interest deductions creates a different incentive structure (Bourassa and Hoesli, 2010) \cite{bourassa2010swiss}. 

Mortgage market structures, interest rate levels, and tax policies differ substantially across developed economies, generating heterogeneous conditions for household leverage strategies. 
Calza et al. (2013) \cite{calza2013housing} analyse how cross-country differences in mortgage market institutions affect the transmission of monetary policy to household consumption. They find that countries with greater reliance on variable rate mortgages and easier refinancing options exhibit stronger consumption responses to interest rate changes, as households' cash flows adjust more rapidly to policy shifts. For debt recycling strategies, these institutional characteristics matter because they determine the volatility of debt service burdens and the flexibility to adjust positions as market conditions evolve.
The real estate finance literature emphasizes that housing market dynamics also vary internationally due to differences in land use regulation, construction elasticity, and demographic trends (Hilber and Vermeulen, 2016) \cite{hilber2016impact}.

\section{Methods}\label{sec: methods}
Our methodological approach builds on the base model \cite{aufiero2025phase} by incorporating mortgage interest rate $r_m$, borrowing interest rate $r_b$ on equity-backed loans, and marginal tax rates $\tau_m$ and $\tau_b$ that create tax shields on these interest payments. These additions modify the dynamical equations that govern the evolution of equity and mortgage at each time step. We retain the analytical tractability of the base model, deriving closed-form expressions for the average processes by computing the eigenvalues and eigenvectors of the modified average transition matrix.

\subsection{Base Model: Equity and Mortgage Dynamics}
We summarise here the foundational debt recycling model developed by Aufiero et al. (2025) \cite{aufiero2025phase}.

The model tracks the joint evolution of two variables in discrete time $t = 0, \ldots, T$: equity $E_t$ (the value of the house portion owned by the borrower) and mortgage balance $M_t$ (the remaining debt). At any time, the house value $H_t$ satisfies:

\begin{equation}
H_t = E_t + M_t \ .
\end{equation}

The usable equity (i.e., the amount that can be leveraged to back additional borrowing) is determined by the loan-to-value (LTV) ratio $\ell \in [0,1]$:

\begin{equation}
U_t = \ell E_t \ .
\end{equation}

The LTV ratio -- calculated by dividing the amount of the loan by the
appraised value or purchase price of the property -- is used by financial institutions and other types of lenders to assess the lending risk before approving a mortgage \cite{brueggeman2011real}. Typically, loan assessments with high ($>80\%$) LTV ratios are considered higher-risk. 
LTV ratios in the UK are typically in the range of $60\%$ to $90\%$, with very few loans with an LTV up to the maximum of $95\%$ available. Similarly, the US is also rather risk averse, with LTV generally capped at $80\%$ and private mortgage insurance being generally required if the LTV ratio goes above that. New Zealand is even more risk averse, with very few loans with LTV ratios above $80\%$. In contrast, Australia is very risk-prone, and regularly allows and issues loans with LTV ratios of $100\%$ and over \cite{lang2020trends, aufiero2025phase}. This creates an economic climate ripe for debt recycling, as shown by our analysis, and may help explain why Australia is currently the only country where this strategy is seen as generally viable and regularly oﬀered as an option.

At each time step, a fraction of the usable equity is invested in a risky asset. The investment return $I_t$ depends on both the risk factor $\mu \in [0,1]$ and a random outcome variable $\sigma_t$:

\begin{equation}
I_t = \sigma_t \mu U_t \ ,
\end{equation}

where $\sigma_t \in \{-1, +1\}$ represents investment success ($+1$) or failure ($-1$). The random variables $\{\sigma_0, \ldots, \sigma_T\}$ are independent Bernoulli draws with probability $p$ of success:

\begin{equation}
p(\sigma) = p\delta_{\sigma,+1} + (1-p)\delta_{\sigma,-1} \ .
\end{equation}

The expected investment outcome is therefore $\langle \sigma \rangle = 2p - 1$ \ .
The housing market also fluctuates, affecting property value at each time step by an amount $H_{t-1}r_t$, where $r_t$ represents the percentage price change. The market fluctuations are assumed normally distributed:

\begin{equation}
p(r) \sim \mathcal{N}(s, \phi^2) \ , \end{equation}
with mean $s$ and variance $\phi^2$. Positive $s$ indicates market's growth; negative $s$ indicates a decline.
The borrower makes scheduled mortgage payments $\pi_t$ at each time step. These payments are assumed non-negative, with occasional skipped payments occurring with a small probability $q \ll 1$. Hence, $p(\pi)$
can be written as
\begin{equation}
    p(\pi) = q\delta(\pi) + (1-q)\delta(\pi - \pi^\star)\ , 
 \end{equation}
yielding an average payment $\langle \pi \rangle = (1-q)\pi^\star$.

The equity and mortgage evolve according to the coupled system of equations

\begin{align}
E_t &= E_{t-1} + \pi_t + I_{t-1} + H_{t-1}r_t \ ,  \\ M_t &= M_{t-1} - \pi_t - I_{t-1} \ . \label{eq:modeloriginal}
\end{align}

Equity increases through scheduled payments $\pi_t$, successful investments (i.e., positive $I_{t-1}$), and appreciation of the housing market (i.e., positive $r_t$). The mortgage decreases symmetrically due to repayments  and positive returns on the  investment that are used to repay the mortgage as well. It increases when investments are not successful, i.e., they lead to money loss.
Through algebraic manipulation, the system can be rewritten as:

\begin{align}
E_t &= \alpha_t E_{t-1} + \pi_t + r_t M_{t-1} \ , \\ M_t &= M_{t-1} - \pi_t + \delta_t E_{t-1}\ ,
\end{align}
where

\begin{align}
\alpha_t &= 1 + \ell\mu\sigma_{t-1} + r_t, \\ \delta_t &= -\ell\mu\sigma_{t-1}. 
\end{align}

In matrix form:

\begin{equation}
\begin{pmatrix} E_t \\ M_t \end{pmatrix} = \begin{pmatrix} \alpha_t & r_t \\ \delta_t & 1 \end{pmatrix} \begin{pmatrix} E_{t-1} \\ M_{t-1} \end{pmatrix} + \begin{pmatrix} \pi_t \\ -\pi_t \end{pmatrix}. 
\end{equation}

The process terminates at the first hitting time $t^\star = \min(t_E, t_M)$, where $t_E$ is the first time equity reaches zero ($E_{t_E} = 0$) and $t_M$ is the first time the mortgage is fully repaid ($M_{t_M} = 0$).
Three outcomes are possible:
\begin{itemize}
    \item Default: If $t^\star = t_E$, equity is depleted before mortgage repayment, signaling strategy failure.
    \item Success: If $t^\star = t_M$, the mortgage is fully repaid. Success is classified as strong success if $t_M < M_0/\pi^\star$ (faster than standard repayment), or weak success if $t_M > M_0/\pi^\star$ (slower than standard repayment).
    \item Permanent re-mortgaging: Neither boundary is reached and continuous refinancing occurs indefinitely.
\end{itemize}

To characterize outcomes analytically, Aufiero et al. (2025) \cite{aufiero2025phase} focus on the average processes $\langle E_t \rangle$ and $\langle M_t \rangle$. The average evolution matrix is:

\begin{equation}
\bar{A} = \begin{pmatrix} 1 + \ell\mu(2p-1) + s & s \\ -\ell\mu(2p-1) & 1 \end{pmatrix} \ ,
\end{equation}
with eigenvalues:

\begin{equation}
\lambda_1 = s + 1, \quad \lambda_2 = \ell\mu(2p-1) + 1\ .
\end{equation}

The eigenvalues determine the phase structure. Through matrix decomposition $\bar{A} = U\Lambda U^{-1}$, the average processes can be expressed as:

\begin{align}
\frac{\left\langle E_t\right\rangle}{\left\langle E_0\right\rangle}  &= \mathcal{A}\lambda_1^t + \mathcal{B}\lambda_2^t + \mathcal{C}, \label{eq:avg_Et} \\
\frac{\left\langle M_t\right\rangle}{\left\langle M_0\right\rangle}  &= \mathcal{D}\lambda_1^t + \mathcal{E}\lambda_2^t + \mathcal{F}, \label{eq:avg_Mt}
\end{align}
where the coefficients $\{\mathcal{A}, \mathcal{B}, \mathcal{C}, \mathcal{D}, \mathcal{E}, \mathcal{F}\}$ depend on initial conditions, payment schedules, and model's parameters (explicit expressions provided in \cite{aufiero2025phase}).

\subsection{Modelling Mortgage Interests and Tax Shield Effects}

We augment the base model to incorporate two new features: (i) interest costs on borrowing and (ii) their associated tax shields. These additions modify the fundamental dynamics, while preserving the analytical tractability of the average process framework.
In each period, the borrower incurs interest charges on both the outstanding mortgage balance and the equity-backed credit line used for investment. In addition, tax deductions on these interest payments create shields that partially offset borrowing costs. Let $r_b \geq 0$ denote the per-period interest rate on the investment credit line, $r_m \geq 0$ the mortgage interest rate, $\tau_b \in [0,1]$ the marginal tax rate applicable to investment interest deductions, and $\tau_m \in [0,1]$ the marginal tax rate for mortgage interest deductions (often zero for owner-occupied properties, but positive for rental properties in many jurisdictions). We modify the original dynamical model in Eq.~\ref{eq:modeloriginal} to incorporate these effects, yielding:

\begin{align}
E_t &= E_{t-1} + \pi_t + I_{t-1} + H_{t-1}~r_t - (1-\tau_b)~r_b ~U_{t-1} + \tau_m r_m M_{t-1}, \label{eq:avg_Et2} \\
M_t &= (1 + r_m)M_{t-1} - \pi_t - I_{t-1}, \label{eq:avg_Mt2}
\end{align}
where $U_{t-1} = \ell E_{t-1}$ and $I_{t-1} = \sigma_{t-1}\mu U_{t-1}$ as before.

The borrowing cost $ (1-\tau_b)~ r_b ~ U_{t-1}$, or equivalently $(1-\tau_b)~r_b~\ell E_{t-1}$, represents the after-tax interest paid on the equity-backed loan, which reduces the available equity. Without the tax shield ($\tau_b = 0$), the full interest amount $r_b \ell E_{t-1}$ is paid. The tax shield reduces this burden proportionally to the marginal tax rate.
The mortgage tax benefit term $ \tau_m r_m M_{t-1} $ represents the tax savings derived from deducting mortgage interest, which flows back into equity. This term is positive only when mortgage interest is tax-deductible (typically for income-producing properties); for owner-occupied residences, $\tau_m = 0$ in most jurisdictions.
The mortgage interest accrual term $(1 + r_m)M_{t-1}$ captures the growth of outstanding mortgage debt due to interest charges before the period's payment and investment transfer are applied.

Taking the expectations over the random variables $\sigma, r, \pi$ using $\langle \sigma_t \rangle = 2p-1$, $\langle r_t \rangle = s$, and $\langle \pi_t \rangle = (1-q)\pi^\star =: \langle \pi \rangle$, the average evolution matrix becomes:

\begin{equation}\label{eq:matrix finale}
\tilde{A} = \begin{pmatrix} 1 + s + \ell\mu(2p-1) - \ell(1-\tau_b)r_b & s + \tau_m r_m \\ -\ell\mu(2p-1) & 1 + r_m \end{pmatrix}\ , 
\end{equation}
where the eigenvalues of the evolution matrix $\tilde{A}$ are the roots of the quadratic: 
\begin{equation}\label{eq:M5_charpoly}
\lambda^2 - \operatorname{tr}\!\big(\tilde{A} \big)\,\lambda + \det\!\big(\tilde{A}\big)=0,
\end{equation}
with

\begin{align}
\operatorname{tr}\!\big(\tilde{A}\big) \;&=\; \big[ 1+s+\ell\mu(2p-1)-\ell(1-\tau_b)r_b\big] \;+\; (1+r_m)\ ,\label{eq:M5_trace}\\
\det\!\big(\tilde{A} \big) \;&=\; \big[1+s+\ell\mu(2p-1)-\ell(1-\tau_b)r_b\big](1+r_m) \;+\; \ell\mu(2p-1)\,\big(s+\tau_m r_m\big)\ .\label{eq:M5_det}
\end{align}

Thus
\begin{equation}
\tilde{\lambda}_{1,2}
=
\frac{1}{2}\left[
\operatorname{tr}(\tilde{A})
\pm
\sqrt{
\operatorname{tr}(\tilde{A})^2
- 4\,\det(\tilde{A})
}
\right]\ .
\end{equation}

Using the new eigenvalues $\tilde {\lambda}_{1,2}$, the new average process can be expressed as: 

\begin{align}
\frac{\left\langle E_t\right\rangle}{\left\langle E_0\right\rangle} &= \tilde{\mathcal{A}} \tilde{\lambda_1} ^t + \tilde{\mathcal{B}} \tilde{\lambda_2} ^t + \tilde{\mathcal{C}}\ ,\label{eq:avg_Et_new} \\
\frac{\left\langle M_t\right\rangle}{\left\langle M_0\right\rangle} &= \tilde{\mathcal{D}} \tilde{\lambda_1} ^t + \tilde{\mathcal{E}} \tilde{\lambda_2} ^t + \tilde{\mathcal{F}}\ , \label{eq:avg_Mt_new} 
\end{align}

where the constants are given explicitly as

$$
\begin{aligned}
\tilde{\mathcal{A}} & =\frac{\left(c_1+1\right)\left(\tilde{\lambda_1}-1\right)}{\tilde{\lambda_1}-\tilde{\lambda_2}} \, \\
\tilde{\mathcal{B}} & = \frac{c_2\left(\tilde{\lambda_1} -1 \right)-\left(\tilde{\lambda_2} -1\right)\left[c_1\left(\tilde{\lambda_1} -1\right)+c_2+\tilde{\lambda_2} -1\right]}{\left(\tilde{\lambda_1} -\tilde{\lambda_2} \right)\left(\tilde{\lambda_2}-1\right)}\ , \\
\tilde{\mathcal{C}} & = -\frac{c_2}{\tilde{\lambda_2}-1} \ , \\
\tilde{\mathcal{D}} & = -\frac{\left(c_3+1\right)\left(\tilde{\lambda_2}-1\right)}{\tilde{\lambda_1}-\tilde{\lambda_2}}\ , \\
\tilde{\mathcal{E}} & = \frac{-c_4\left( \tilde{\lambda_1}-1\right)+\left( \tilde{\lambda_2} -1\right)\left[c_3\left( \tilde{\lambda_2} -1\right)+ c_4+ \tilde{\lambda_1} -1\right]}{\left( \tilde{\lambda_1} -\tilde{\lambda_2}\right)\left(\tilde{\lambda_2}-1\right)}, \\
\tilde{\mathcal{F}} & = \frac{c_4}{\tilde{\lambda_2}-1}\ ,
\end{aligned}
$$

with

\begin{equation}
c_1=\frac{\left\langle M_0\right\rangle}{\left\langle E_0\right\rangle} \ ; \quad c_2=\frac{\langle\pi\rangle}{\left\langle E_0\right\rangle}\ ; \quad  c_3=\frac{1}{c_1}=\frac{\left\langle E_0\right\rangle}{\left\langle M_0\right\rangle}\ ; \quad c_4=\frac{\langle\pi\rangle}{\left\langle M_0\right\rangle}\ .
\end{equation}

\subsection{Model's  Empirical Calibration}
To ground our tax-augmented debt-recycling model in realistic institutional environments, we calibrate the parameters $(s, r_m, r_b, \tau_m, \tau_b)$ to the case of three jurisdictions: Australia, Germany, and Switzerland. These countries exhibit markedly different mortgage market structures and tax treatment of housing-related interest, providing a representative range of conditions under which debt recycling strategies operate.

\textbf{Australia} represents a market where debt recycling is already an established practice. Variable-rate mortgages dominate, and interest on investment-related borrowing is broadly tax-deductible. Both mortgage rates and borrowing costs are relatively high compared to other developed markets, but substantial tax shields are available, creating conditions where the strategy may be viable despite high interest costs \cite{ATO2025a}.
\textbf{Germany} provides a contrasting regime. Mortgage interest on owner-occupied properties receives no tax benefit, and interest deductibility for investment borrowing is severely restricted beyond a modest saver's allowance. Fixed-rate mortgages are common and rates are moderate \cite{PwC2025GermanyDeductions}. The absence of tax shields for most households means that interest costs exert their full negative effect on debt recycling dynamics.
\textbf{Switzerland} combines low interest rates with a favorable tax regime. Mortgage rates are among the lowest internationally, and the Swiss tax code permits broad deductibility of private debt interest \cite{SNB2025}. For owner-occupied homes, imputed rental income creates taxable income against which interest can be deducted, effectively providing a tax shield in both ownership scenarios.

Tax treatment varies not only across countries, but also by property use within a jurisdiction. We therefore consider two cases for each country: \textbf{owner-occupied housing} (if the borrower resides in the property as their primary residence) or \textbf{rental housing} (if the property generates rental income from tenants).
Tab. \ref{tab:calibration} summarises the calibrated parameters for all three countries under both ownership scenarios. Parameter selection is based on official government sources, central bank data, and statutory tax codes current as of 2024-2025
\cite{ATO2025a, RBA2025b, RBA2025a, ABS2021, PwC2025GermanyDeductions, BMJ2025, Destatis2025, Fedlex2025, ESTV2024, SNB2025}.

The main calibrated parameters are reported in Table \ref{tab:calibration}.

\begin{table}[h!]
\centering
\caption{Country parameters by ownership type (2024-2025).}
\label{tab:calibration}
\footnotesize
\begin{tabular}{|l|l|l|l|l|l|l|}
\hline
 & \multicolumn{2}{|c|}{Australia} & \multicolumn{2}{|c|}{Germany} & \multicolumn{2}{|c|}{Switzerland} \\
\hline
 & Self-Owned & Rental & Self-Owned & Rental & Self-Owned & Rental \\
\hline
$s$ & $[-4\%, 4\%]$ & $[-4\%, 4\%]$ & $[-6\%, 6\%]$ & $[-6\%, 6\%]$ & $[-2\%, 2\%]$ & $[-2\%, 2\%]$ \\
\hline
$r_b$ & 8.35\% & 8.35\% & 5.34\% & 5.34\% & 1.62\% & 1.62\% \\
\hline
$r_m$ & 6.07\% & 6.28\% & 3.69\% & 3.69\% & 1.62\% & 1.62\% \\
\hline
$\tau_b$ & 32\% & 32\% & 0\% & 0\% & 20.1\% & 20.1\% \\
\hline
$\tau_m$ & 0\% & 32\% & 0\% & 37.5\% & 20.1\% & 20.1\% \\
\hline
\end{tabular}
\end{table}

Tax brackets are selected based on the average national income to represent typical taxpayers. Australia's Average Weekly Ordinary Time Earnings, Germany's average gross employee earnings, and Switzerland's national wage statistics are converted to GBP at a common exchange rate snapshot for comparability. Country-specific tax calculators and statutory formulas yield the marginal rates reported.
Published annual rates $(r_m^{\text{annual}}, r_b^{\text{annual}})$ are converted to quarterly equivalents to match the model's discrete time steps, i.e.,  $r_m = (1 + r_m^{\text{annual}})^{1/4} - 1$, and similarly for $r_b$.

In this model, we assume the following: (i) investment credit lines fund only income-producing assets (no personal consumption); (ii) rental properties maintain continuous occupancy (no vacancy periods); (iii) for rental properties, the rental income equals and thereby covers the regular installments $\pi^\star$, hence abstracting from explicit modeling of rental cash flows; (iv) housing market shift ranges apply uniformly to owner-occupied and rental properties within each country, as national indices do not distinguish by use; (v) capital gains taxes on investment returns are omitted; (vi) mortgage interest each quarter is computed on the opening balance $r_m M_{t-1}$, with net principal change $\Delta M_t = -(\pi_t + I_{t-1} - r_m M_{t-1})$.

\section{Results}\label{sec: Results}

This section presents the outcomes of the tax-augmented model calibrated to the country-specific parameters (see Tab. \ref{tab:2025_calibration}).
In the following scenarios, the parameters of the base model are set at $\ell = 0.5$, $\mu =0.5$, $E_0 = \$ 30,000$, $M_0 = \$ 300, 000$, $\pi^\star = \$3,000$, $q = 1\%$, while we analyse the effect of the calibrate tax and interest cost rates $(s, r_m, r_b, \tau_m, \tau_b)$ for different home-ownership scenarios.

\subsection{Owner-Occupied Housing}

\subsubsection{Phase Diagrams}

\begin{figure}[ht]
    \centering

    \begin{subfigure}[b]{0.49\textwidth}
        \centering
        \includegraphics[width=\textwidth]{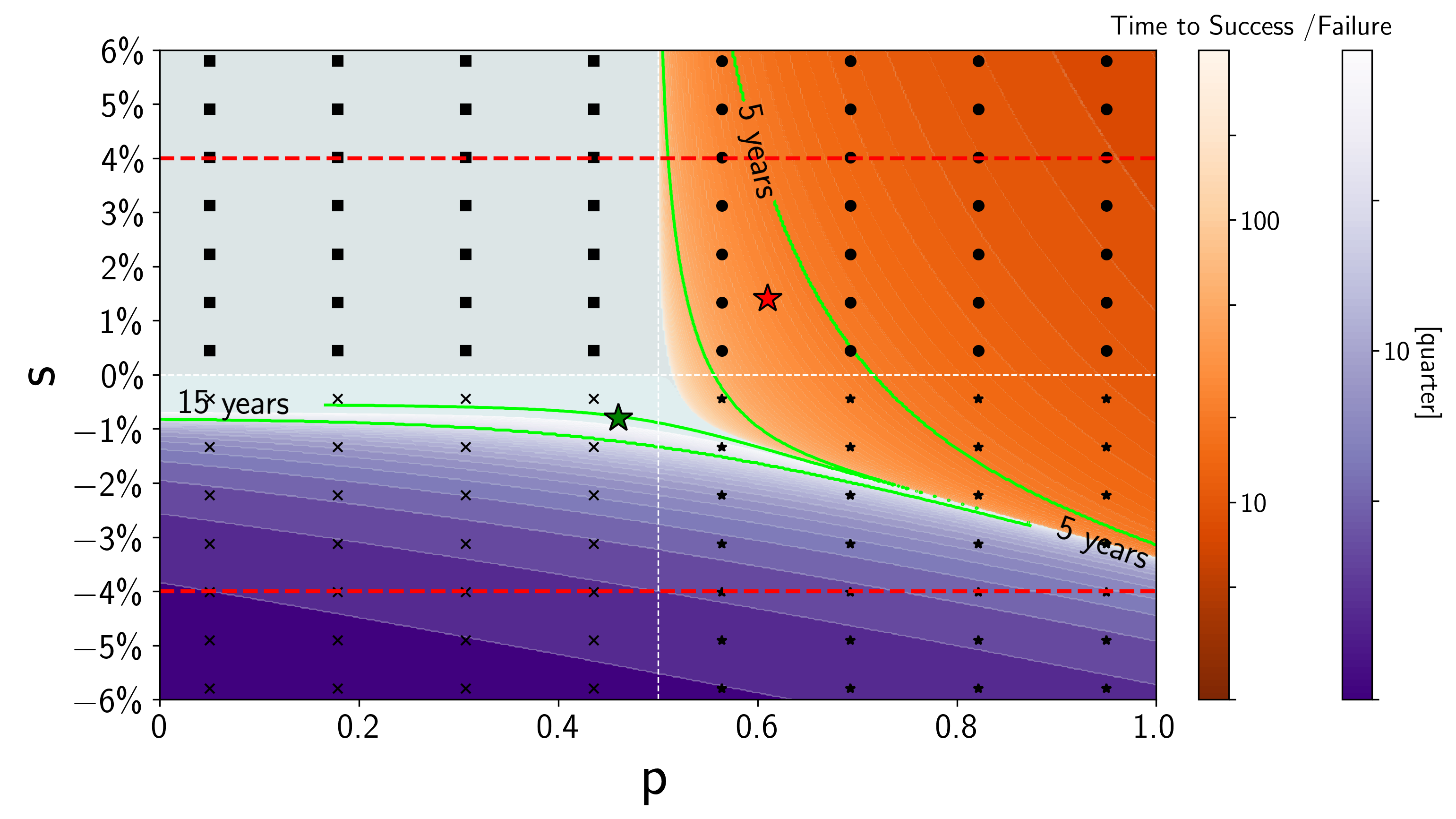}
        \caption{Australia}
        \label{fig:australia_owned}
    \end{subfigure}
    \begin{subfigure}[b]{0.49\textwidth}
        \centering
        \includegraphics[width=\textwidth]{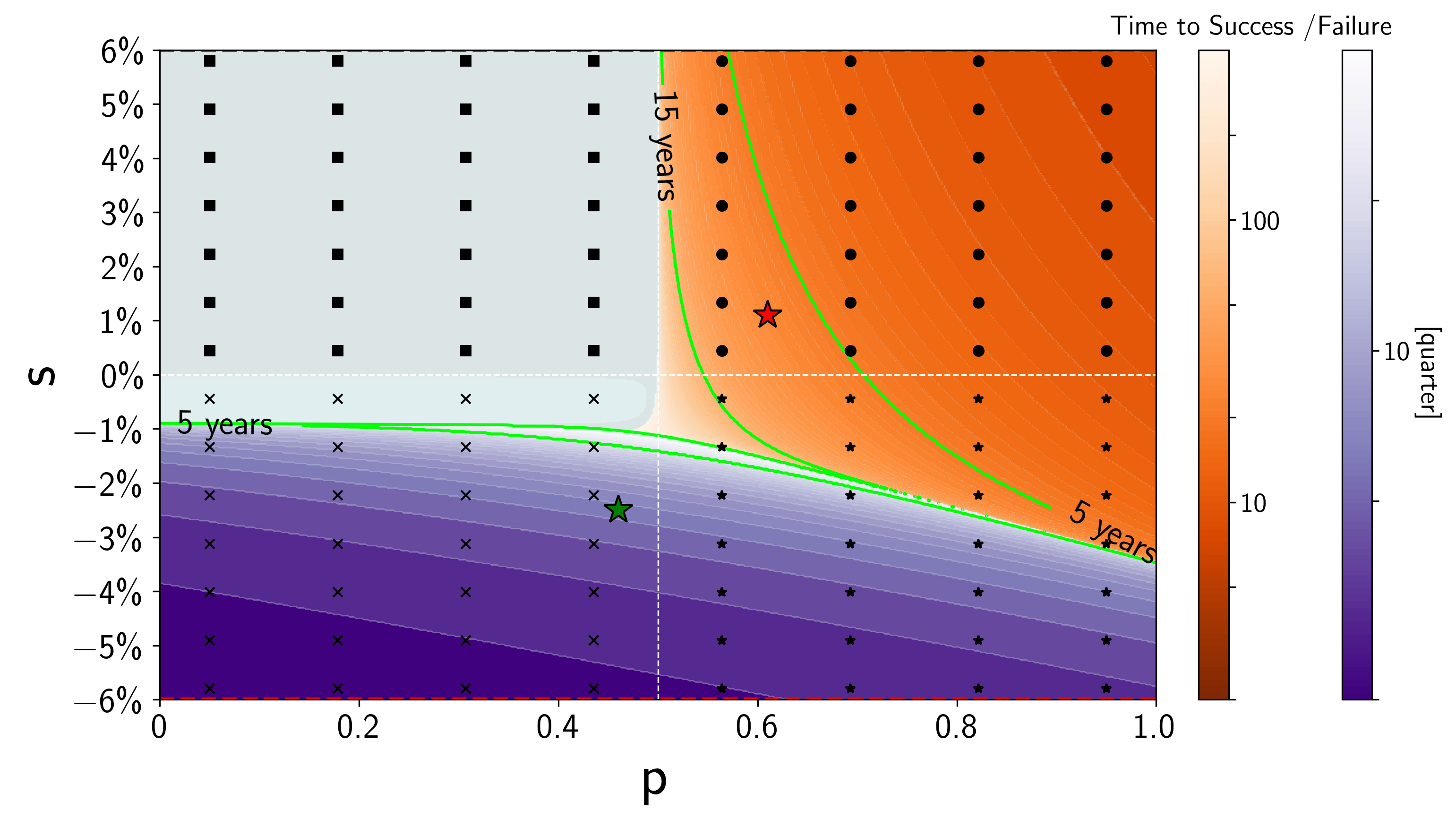}
        \caption{Germany}
        \label{fig:germany_owned}
    \end{subfigure}

    \begin{subfigure}[b]{0.49\textwidth}
        \centering
        \includegraphics[width=\textwidth]{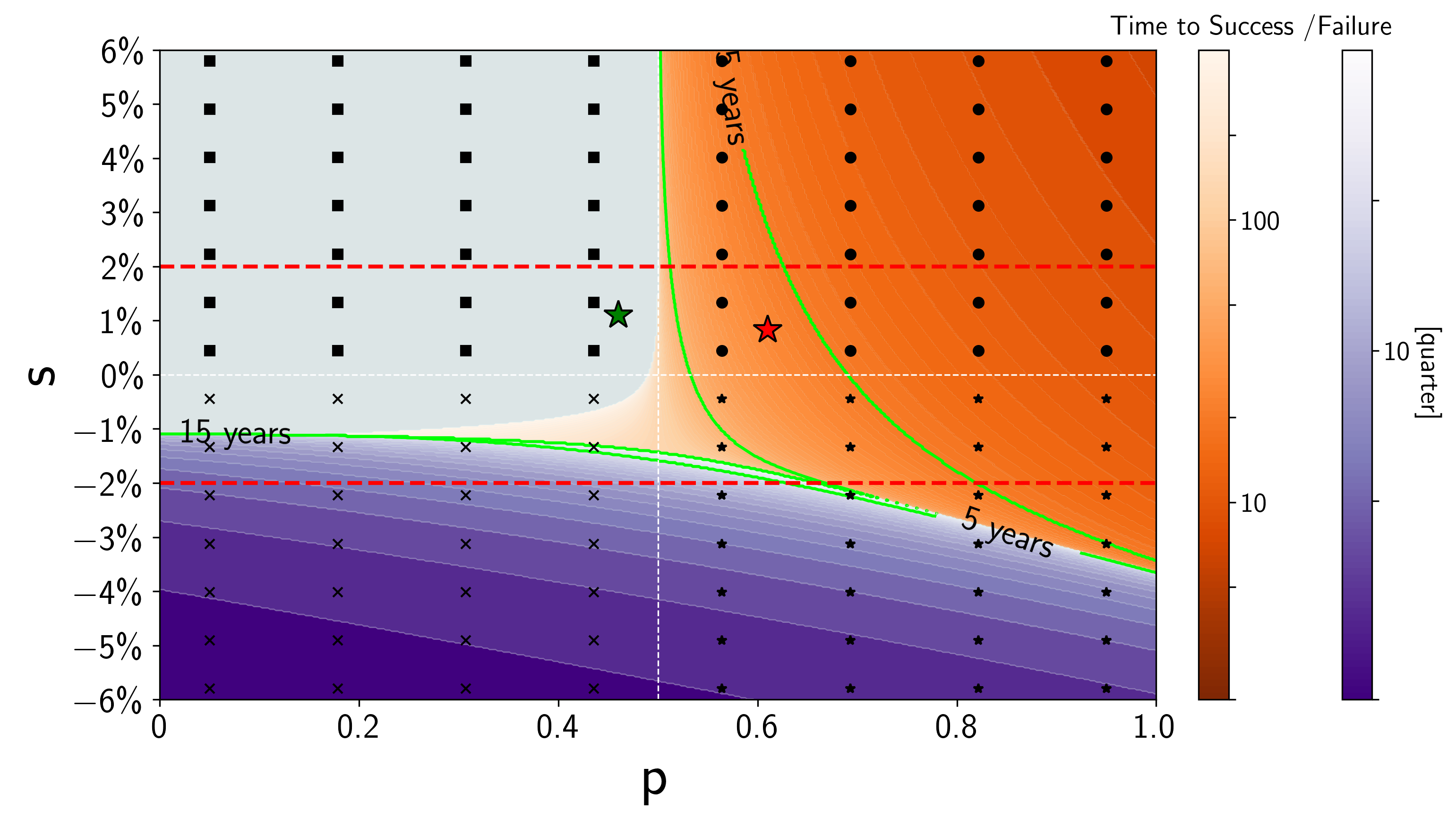}
        \caption{Switzerland}
        \label{fig:switzer_owned}
    \end{subfigure}
    \caption{Calibrated Phase Diagrams for the Self-Owned Case in Australia, Germany and Switzerland with parameters $\ell = 0.5$, $\mu =0.5$, $E_0 = \$ 30,000$, $M_0 = \$ 300, 000$, $\pi^\star = \$3,000$, $q = 1\%$ and $(s, r_m, r_b, \tau_m, \tau_b)$ specified in Tab. \ref{tab:calibration}. Stars represent market conditions (housing-investment) in two representative years, namely Q4 2008 (financial crises, green) and Q2 2025 (current, red). The red dashed lines represent the range of the typical housing market fluctuations ($s$) observed in the jurisdiction.}
    \label{fig: self owned PD}
\end{figure}

Fig. \ref{fig: self owned PD} presents phase diagrams for owner-occupied properties calibrated to tax and interests parameters for Australia, Germany, and Switzerland. Orange regions indicate debt recycling success outcome (i.e., the mortgage is repaid before equity depletion), with color intensity proportional to the speed at which we reach the outcome. Purple regions indicate default (i.e., the equity is depleted before mortgage repayment). Gray regions denote permanent re-mortgaging, where neither absorbing boundary ($E=0$ or $M=0$) is reached within the $400$-quarters horizon. Lime contour lines mark constant first-hitting times at $5$ and $15$ years. The vertical axis is standardised across all panels to span the widest observed housing-drift range, from $-6\%$ to $+6\%$ (see Tab. \ref{tab:2025_calibration}), to permit easier cross-country comparison. Country-specific drift intervals are then highlighted with red horizontal lines, marking the corresponding upper and lower bounds for Australia, Germany, and Switzerland.

Australia exhibits the smallest success region and the widest default zone for negative $s$. This unfavorable outcome structure follows directly from the combination of high interest rates ($r_m = 6.07\%$, $r_b = 8.35\%$) and the absence of a mortgage tax shield for owner-occupied properties ($\tau_m = 0\%$). The high borrowing cost $r_b U_{t-1}$ substantially reduces equity (see Eq. \ref{eq:avg_Et2}), despite the partial offset from the relatively high investment tax shield ($\tau_b = 32\%$). Without the mortgage tax benefit ($\tau_m r_m M_{t-1} = 0$),  equity receives no cushion from mortgage interest deductibility. The mortgage accrues interest at $r_m = 6.07\%$ in Eq. \ref{eq:avg_Mt2}, creating a substantial hurdle that scheduled payments and investment returns must overcome to achieve principal reduction.

Germany presents a moderately more favorable phase structure than Australia at comparable $(p,s)$ coordinates, despite also offering no tax shields for owner-occupiers ($\tau_m = \tau_b = 0\%$)\footnote{Although the plots do not display points with identical ($p,s$) coordinates, this statement refers to the expected comparison that would emerge if such pairs were selected. A representative ($p,s$) combination and their average trajectories are examined later in Fig. \ref{fig:success_self owned}, where differences in adjustment speed become evident.}. The improvement stems entirely from lower interest rates: $r_m = 3.69\%$ and $r_b = 5.34\%$. 
The reduced borrowing cost slows equity depletion in each period, and the lower mortgage rate increases the likelihood that $\pi_t + I_{t-1}$  exceeds $r_m M_{t-1}$, enabling principal reduction rather than outstanding balance growth. The success region expands relative to Australia, and iso-time contours shift leftward, indicating faster outcomes where success occurs.

Switzerland displays the most extensive success region and the most compressed iso-time contours among the three countries. This favorable configuration results from both low rates ($r_m = r_b = 1.62\%$) and positive tax shields on both borrowing and mortgage interest ($\tau_m = \tau_b = 20.1\%$). 
These combined effects dramatically reduce the interest burden, allowing success even for moderate investment performance ($p \approx 0.5$) and slightly negative housing markets ($s < 0$).

To relate the calibrated model to current conditions, we identify a pair of coordinates $(p, s)$ that capture the 2025 macro-financial environment in each country. This would allow us to determine which region of the phase diagram would correspond to the present situation if the strategy were implemented today. For the calibration of the housing market drift $s$ and the investment success probability $p$, we employ country-specific and market-based estimates corresponding to the second quarter of 2025 (April-June).
Official statistics indicate that the mean quarterly housing appreciation equals $s_A=1.41\%$ for Australia \cite{abs_total_value_of_dwellings_2025}, $s_G=1.10\%$ for Germany \cite{destatis_house_price_index_germany}, and $s_S=0.83\%$ for Switzerland \cite{gpg_switzerland_home_price_trends_2025}.

To represent a realistic benchmark for household investment decisions, we assume the risky asset corresponds to a diversified equity index, consistent with moderate-risk, long-term investment behavior. The probability of investment success $p$ is therefore calibrated from recent S\&P 500 data: $p$ is computed as the fraction of trading days with positive daily returns within the quarter, providing a data-driven measure of the market's short-term momentum over that horizon. Formally, let $r_d$ denote the daily return on the S\&P 500 index and $N$ the number of trading days in the quarter. The empirical success probability is given by
\begin{equation}\label{eq: calibration p}
\hat{p}=\frac{1}{N} \sum_{d=1}^N \textbf{1}\left\{r_d>0\right\},
\end{equation}
where $\textbf{1}\{\cdot\}$ denotes the indicator function, taking the value 1 when the condition is satisfied.
For Q2 2025, S\&P 500 closing prices yield an estimated $p=0.61$ $\big( N=61, N_{+}=37 \big)$; repeating the same procedure for the STOXX Europe 600 index gives $p= 0.54$ $\big( N=59, N_{+}=32 \big)$.
The resulting $(p,s)$ pairs are reported in Tab. \ref{tab:2025_calibration}, and indicated by the red stars in Fig. \ref{fig: self owned PD}. 
The model average dynamics is placed firmly in the first quadrant -- corresponding to positive housing drift ($s>0$) and a fair probability of gaining from the investment ($p>0.5$) -- for all combinations of jurisdictions and indices considered. As observed, in all three countries these coordinates lie within the strong success region, indicating that under current market conditions, pursuing the debt recycling strategy would be advisable.

\begin{table}[h!]
\centering
\caption{Q2 2025.}
\footnotesize
\begin{tabular}{|l|c|c|c|c|}
\hline
\textbf{Country} & \textbf{$s$ Q2 2025} & \textbf{$p$ Q2 2025} & \textbf{$s$ Q4 2008} & \textbf{$p$  Q4 2008}\\
\hline
Australia & $+1.41\%$ & $0.61$ & $-0.8\%$ & $0.46$ \\
\hline
Germany & $+1.10\%$ & $0.61$ & $-2.5\%$ & $0.46$ \\
\hline
Switzerland & $+0.83\%$ & $0.61$ & $+1.1\%$ & $0.46$ \\
\hline
\end{tabular}
\label{tab:2025_calibration}
\end{table}

Finally, we focus on the Global Financial Crisis from the third quarter of 2008 to the second quarter of 2009, which includes the default of major financial institutions, the resulting liquidity contraction, and the transmission of systemic stress to real economic and housing market conditions \cite{aufiero2025mapping}.
This episode is represented in Fig. \ref{fig: self owned PD} by the green star, corresponding to the crisis calibration point derived from the observed housing and equity performance during that interval.
In this case, Australia and Germany are positioned in the third quadrant of the phase diagram, indicating a failure of the strategy (and an especially rapid one in the case of Germany). By contrast, the Swiss configuration lies in the second quadrant, corresponding to the regime of permanent re-mortgaging. These results confirm that, as expected, implementing the strategy during the 2008 episode would have been inadvisable, given the prevailing macro-financial conditions.

\begin{figure}[H]
    \centering

    \begin{subfigure}[b]{0.49\textwidth}
        \centering
        \includegraphics[width=\textwidth]{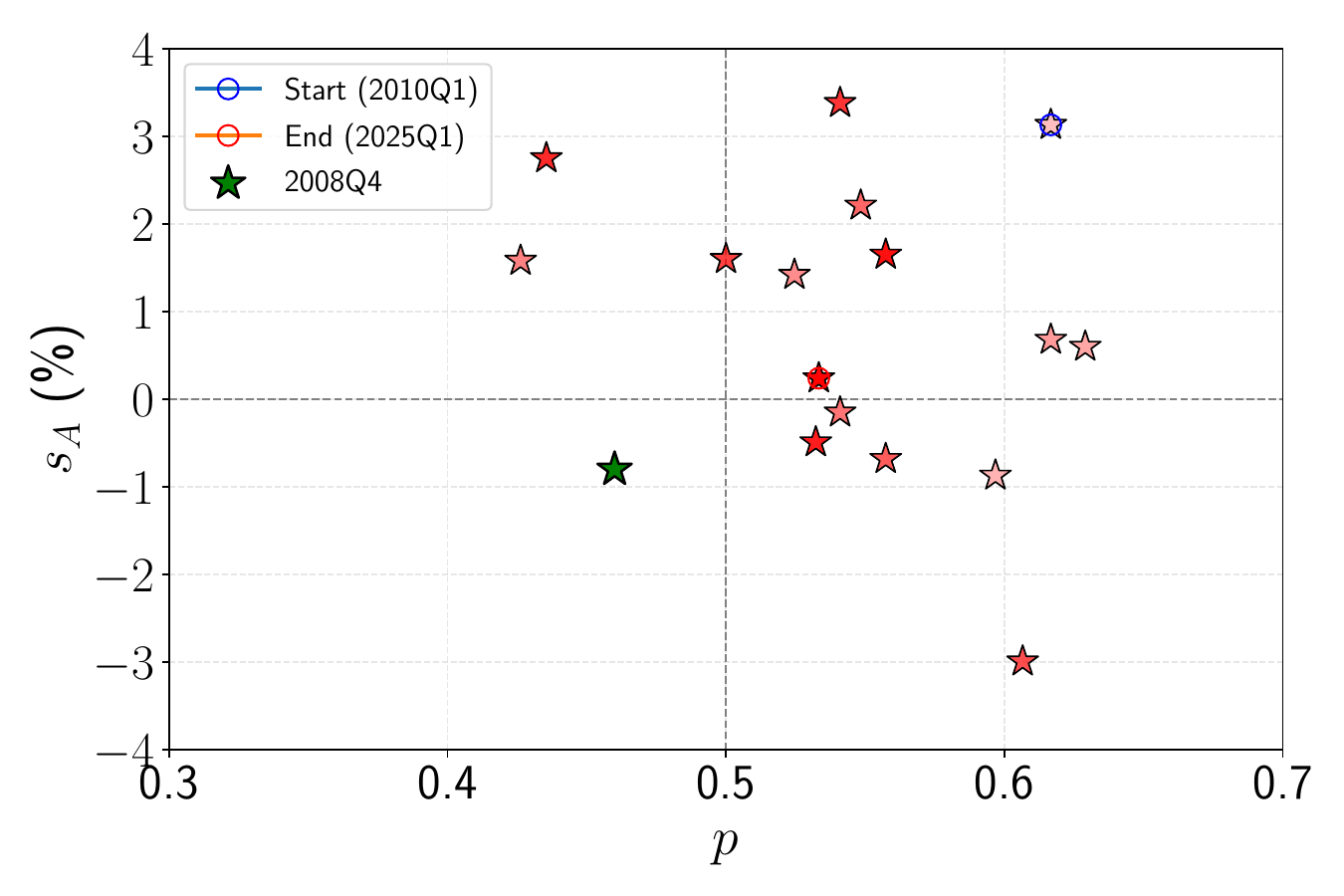}
        \caption{Australia}
        \label{fig:australia time}
    \end{subfigure}
    \begin{subfigure}[b]{0.49\textwidth}
        \centering
        \includegraphics[width=\textwidth]{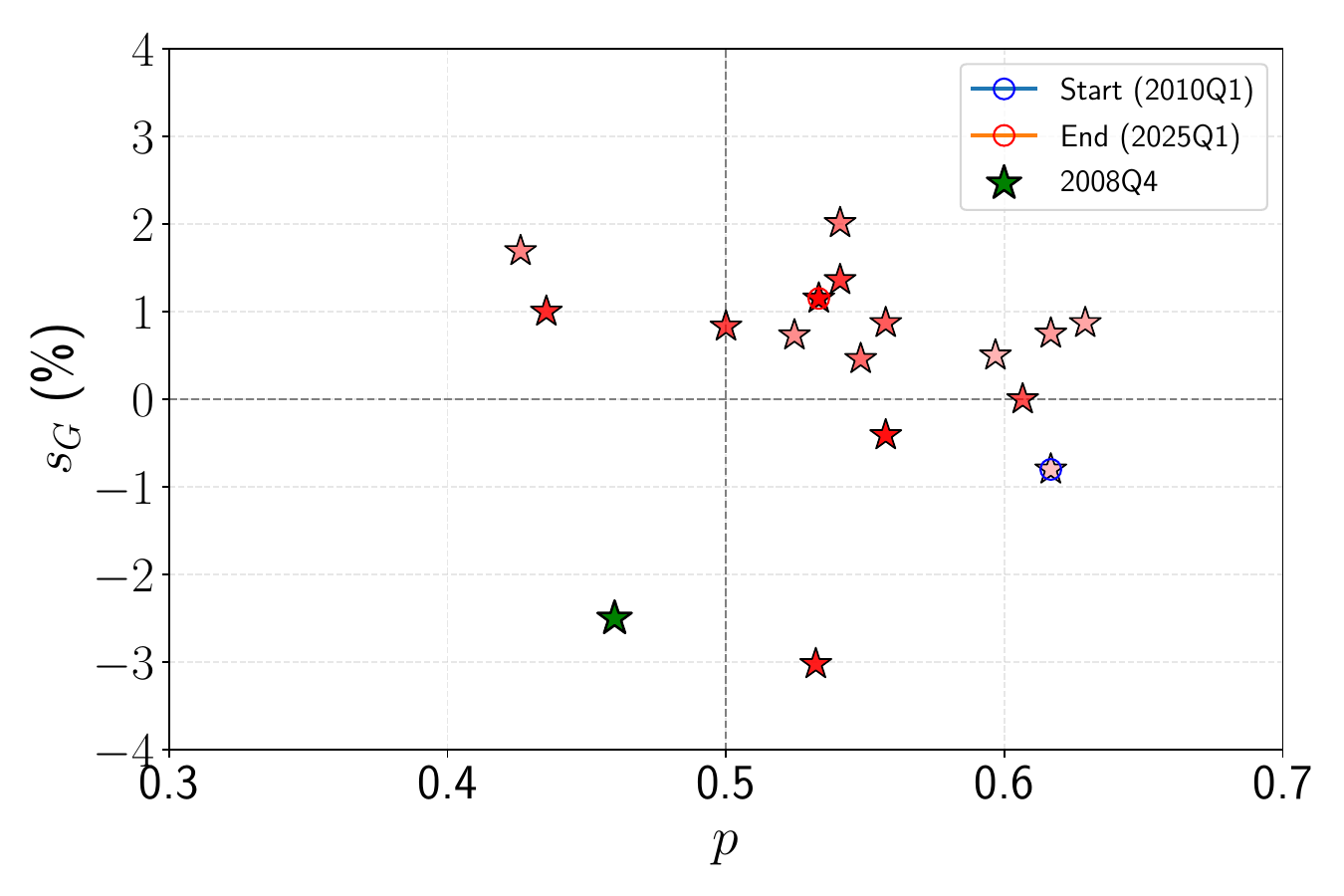}
        \caption{Germany}
        \label{fig:germany time}
    \end{subfigure}

    \begin{subfigure}[b]{0.49\textwidth}
        \centering
        \includegraphics[width=\textwidth]{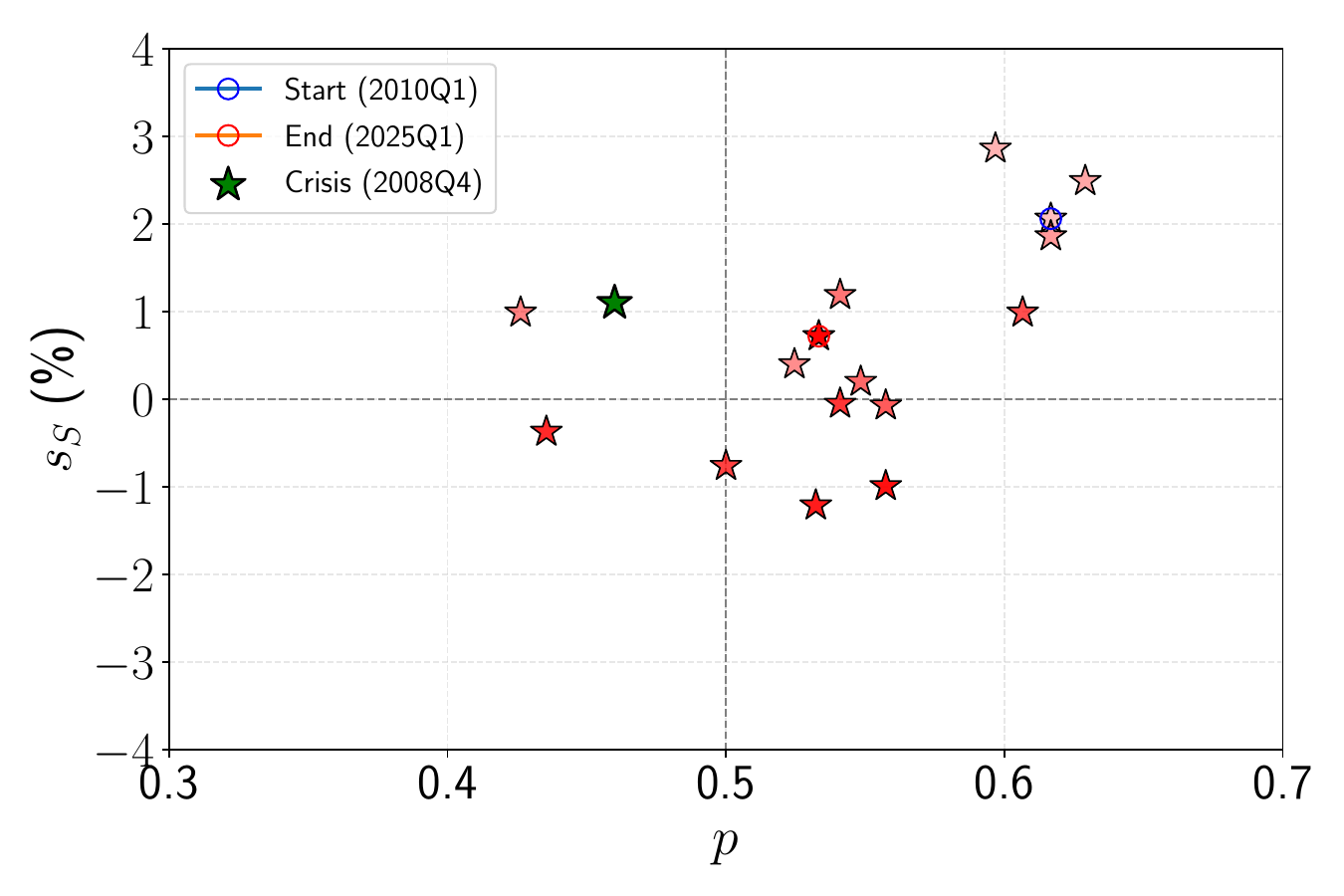}
        \caption{Switzerland}
        \label{fig:switzer time}
    \end{subfigure}
    \caption{Quarterly calibration of ($p,s$) pairs for Australia, Germany, and Switzerland over the period 2010-2025. Each red star corresponds to the first quarter of a given year, with color intensity increasing over time to represent the temporal dimension of the calibration. The blue circle marks the starting point (Q1 2010), while the red circle denotes the most recent observation (Q1 2025). The green star corresponds to the global financial crisis (Q4 2008).}
    \label{fig: calibration time}
\end{figure}
\paragraph{Historical Benchmark}

To ensure a robust and policy-relevant calibration, the probability of investment success $\hat{p}$ is estimated over a $15$ year historical window (Q1 2010 – Q2 2025), aligning with financial planning guidelines suggesting that debt recycling strategies are most suitable for individuals who can sustain the investment cycle for five to fifteen years. For each quarter $q$, we compute $p_q$ as the empirical fraction of trading days with positive daily returns within that quarter (Eq. \ref{eq: calibration p}). 

For $s_q$, we use official quarterly residential property price indices (Q1 2010 – Q2 2025) from national sources harmonised by the BIS Selected Residential Property Prices database (Australia: ABS \cite{BIS_QAUN628BIS_2025}; Germany: Destatis \cite{BIS_QDEN628BIS_2025}; Switzerland: FSO IMPI \cite{BIS_QCHN628BIS_2025}). Quarterly housing drift is the quarter-on-quarter percentage change.

In Fig. \ref{fig: calibration time} we show the calibration of ($p_q, s_q$) over the period 2010-2025. For each country, we plot one red star corresponding to the ($p,s$) pair observed in the first quarter of each year, yielding a total of sixteen points. The color intensity increases progressively from Q1 2010 to Q1 2025, thereby encoding the temporal dimension of the series. The first observation is highlighted with a blue circle, while the most recent one is enclosed in a red circle. Note that the last star differs from those reported in Fig. \ref{fig: self owned PD}, where the marker referred instead to Q2 2025.
Across all jurisdictions, the trajectory remains largely within the first quadrant, highlighting a persistent alignment between moderate housing growth and favourable equity performance throughout the sample period. The dispersion of the stars across quadrants indicates that, in principle, all potential outcomes of the strategy would have been possible in the analysed period.

\subsubsection{Simulations of the Equity and Mortgage Processes}

To illustrate the dynamics underlying the phase diagrams, we conduct Monte Carlo simulations for two representative parameter combinations. The success scenario sets $p = 0.6$ and $s = 1.5\%$, representing favourable investment and housing market conditions. The default scenario sets $p = 0.4$ and $s = -1.5\%$, representing adverse conditions. For each scenario and country, we simulate $N=10^5$ sample paths of the process in Eqs. $\big($\ref{eq:avg_Et2}, \ref{eq:avg_Mt2}$\big)$ using the calibrated parameters in Tab. \ref{tab:calibration}.

\paragraph*{Success Scenario}

\begin{figure}
    \centering
    \includegraphics[width=\linewidth]{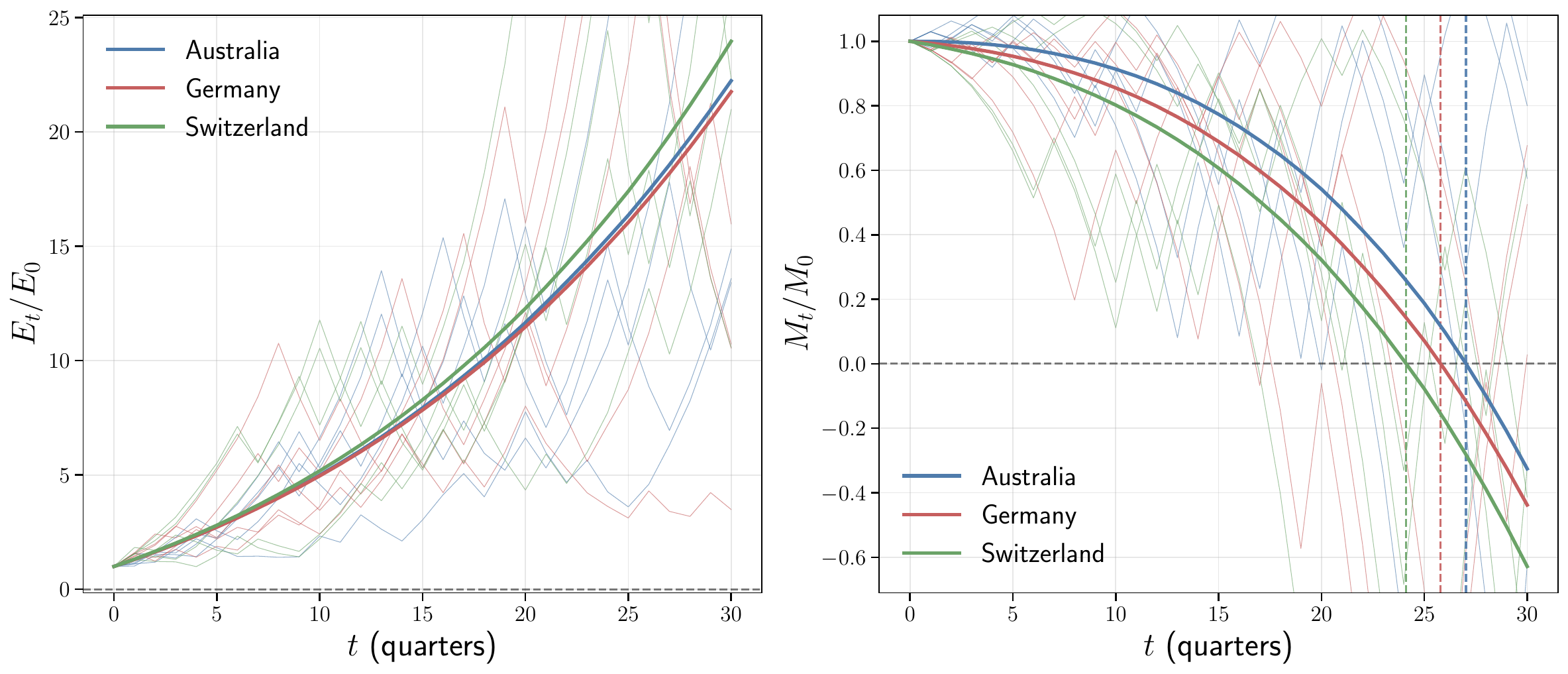}
    \caption{Calibrated Simulation: Success Scenario with  $p = 0.6$ and $s = 1.5\%$  for the  Self-Owned Case.}
    \label{fig:success_self owned}
\end{figure}

Fig. \ref{fig:success_self owned} displays the normalised equity $E_t / E_0$ (left panel) and normalised mortgage $M_t / M_0$ (right panel). The bold line represents the empirical average process, while lighter trajectories illustrate individual simulations. Vertical dashed lines indicate the first-hitting time of the empirical average path: Switzerland is the most favourable (faster) scenario; Germany exhibits slower convergence than Switzerland but faster than Australia -- with their ordering matching what we observed in the phase diagrams (see Fig. \ref{fig: self owned PD}. Australia shows the slowest progression toward mortgage repayment. Without mortgage interest deductibility ($\tau_m = 0\%$), equity grows more slowly than in the other countries, hence lengthening the time to outcome.

\paragraph*{Default Scenario}

\begin{figure}
    \centering
    \includegraphics[width=\linewidth]{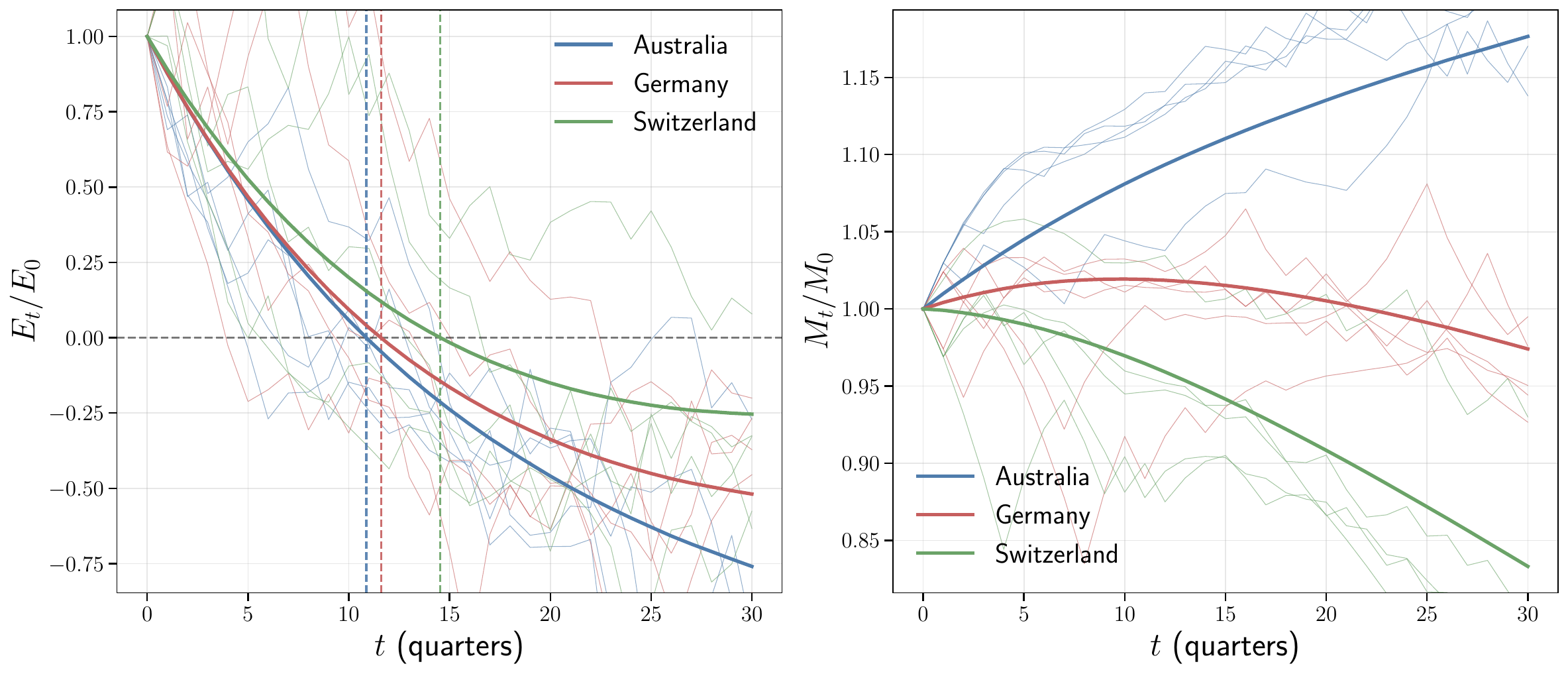}
    \caption{Calibrated Simulation: Default Scenario with  $p = 0.4$ and $s = -1.5\%$ for the Self-Owned Case.}
    \label{fig:default selfowned}
\end{figure}

Fig. \ref{fig:default selfowned} illustrates the default scenario where adverse conditions lead to equity depletion before mortgage repayment. The ordering of first-hitting times reverses: Australia reaches $E_t = 0$ earliest, Germany follows, and Switzerland exhibits the longest survival time before default.
Again, this is consistent with the phase diagrams shown before: the Swiss shield partly offsets the borrowing costs, so the green equity path declines slowest; Germany
lacks shields and therefore sits between Switzerland and Australia: Australia’s higher $r_m$ and $r_b$ push
the effective costs up and equity fails sooner.

\subsection{Rental Properties}

\subsubsection{Phase Diagrams}

\begin{figure}[ht]
    \centering

    \begin{subfigure}[b]{0.49\textwidth}
        \centering
        \includegraphics[width=\textwidth]{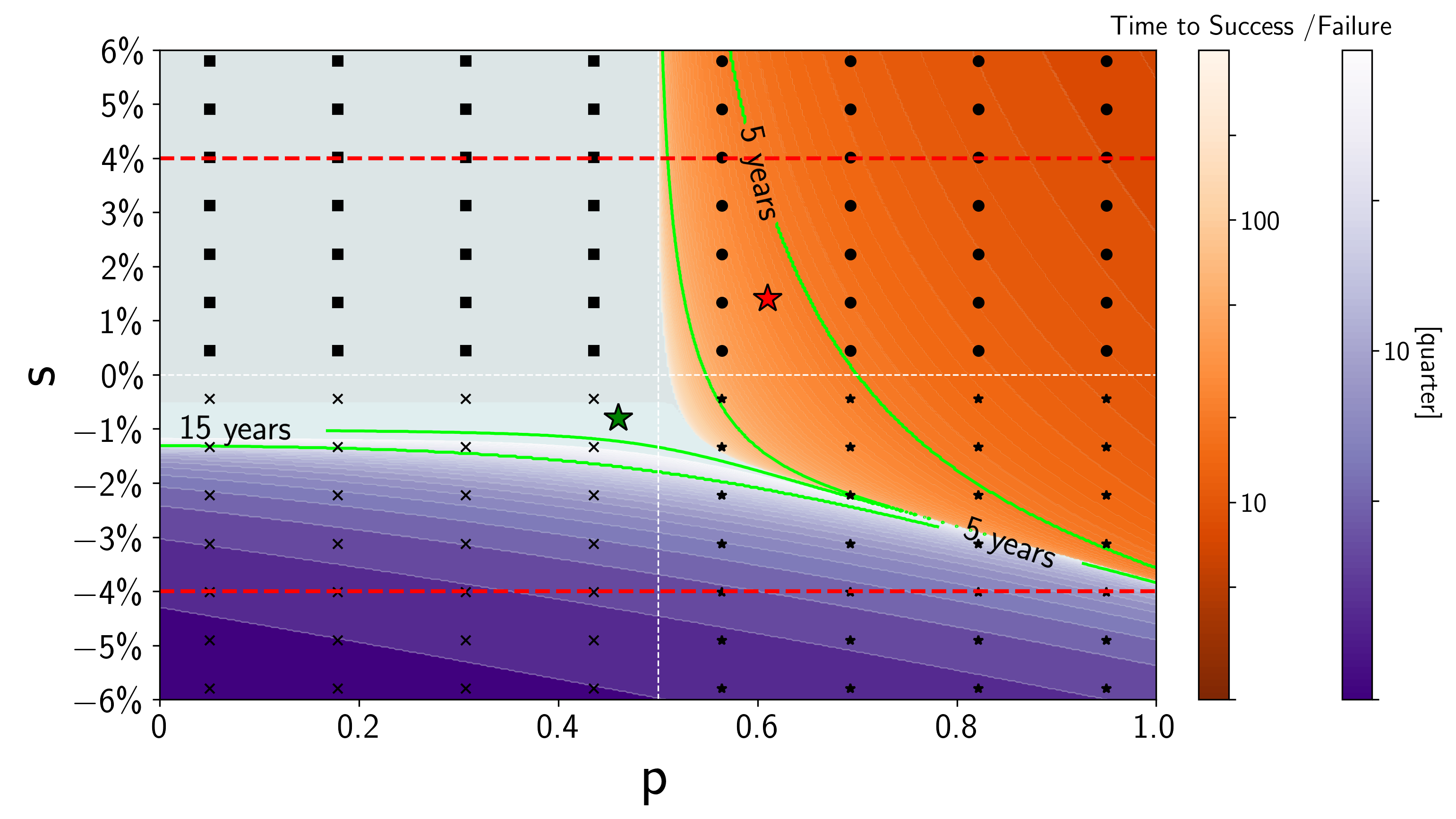}
        \caption{Australia}
        \label{fig:australia_rented}
    \end{subfigure}
    \begin{subfigure}[b]{0.49\textwidth}
        \centering
        \includegraphics[width=\textwidth]{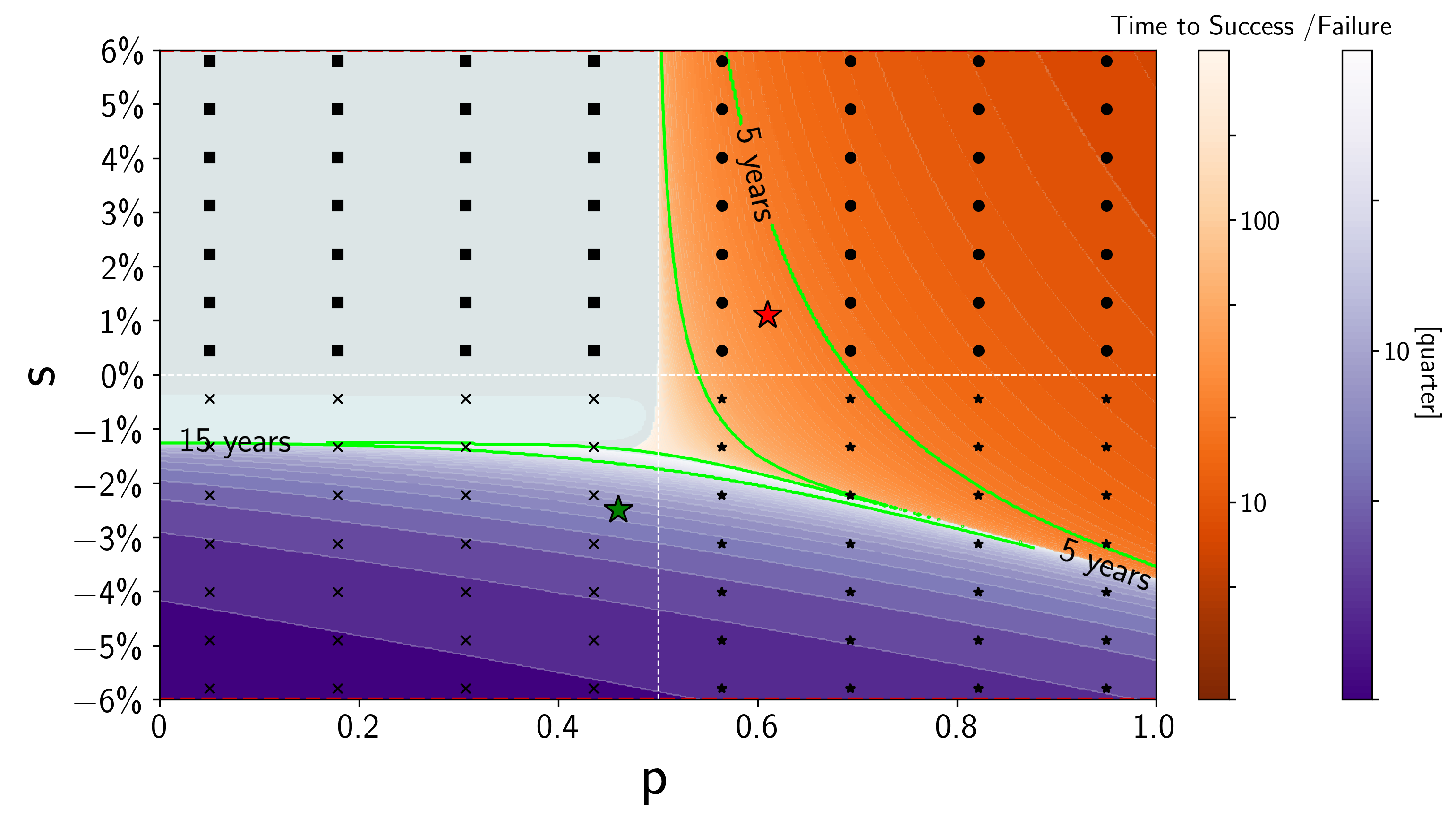}
        \caption{Germany}
        \label{fig:germany_rented}
    \end{subfigure}

    \begin{subfigure}[b]{0.49\textwidth}
        \centering
        \includegraphics[width=\textwidth]{PD_Switzerland_both_-2,2.png}
        \caption{Switzerland}
        \label{fig:switzer_rented}
    \end{subfigure}
    \caption{Calibrated Phase Diagrams for the Rental Case in Australia, Germany and Switzerland with parameters $\ell = 0.5$, $\mu =0.5$, $E_0 = \$ 30,000$, $M_0 = \$ 300, 000$, $\pi^\star = \$3,000$, $q = 1\%$ and $(s, r_m, r_b, \tau_m, \tau_b)$ specified in Tab. \ref{tab:calibration}. Stars represent market conditions (housing-investment) in two representative years, namely Q4 2008 (financial crises, green) and Q2 2025 (current, red). The red dashed lines represent the range of the typical housing market fluctuations ($s$) observed in the jurisdiction.}
    \label{fig: rented PD}
\end{figure}

Fig. \ref{fig: rented PD} presents phase diagrams for rental properties using the calibrated parameters from Tab. \ref{tab:calibration}. Tax treatment differs from the owner-occupied case: mortgage interest is tax-deductible in Australia ($\tau_m = 32\%$) and Germany ($\tau_m = 37.5\%$), while Switzerland maintains the same favourable treatment as for owner-occupiers ($\tau_m = 20.1\%$). Investment borrowing tax shields remain unchanged. Australia and Germany display an expanded success region relative to their owner-occupied counterparts, since the addition of mortgage interest deductibility ($\tau_m >0\%$ versus $0\%$ previously) introduces a boost in equity each period. 
Switzerland remains the most favourable environment among the three, with its phase diagram reproducing the broad success region and compressed timing observed earlier.  The rental parameters are identical to the owner-occupied case ($\tau_m = \tau_b = 20.1\%$, $r_m = r_b = 1.62\%$) due to Switzerland's unified treatment of owner-occupied and rental properties for tax purposes.

\subsubsection{Simulations of the Equity and Mortgage Processes}

We consider here the simulations of the dynamical processes of equity and mortgage using the same benchmark scenarios ($p = 0.6$, $s = 1.5\%$ for success; $p = 0.4$, $s = -1.5\%$ for default) with the rental property tax shield and interest costs parameters in Tab. \ref{tab:calibration}.

\paragraph*{Success Scenario}

\begin{figure}
    \centering
    \includegraphics[width=\linewidth]{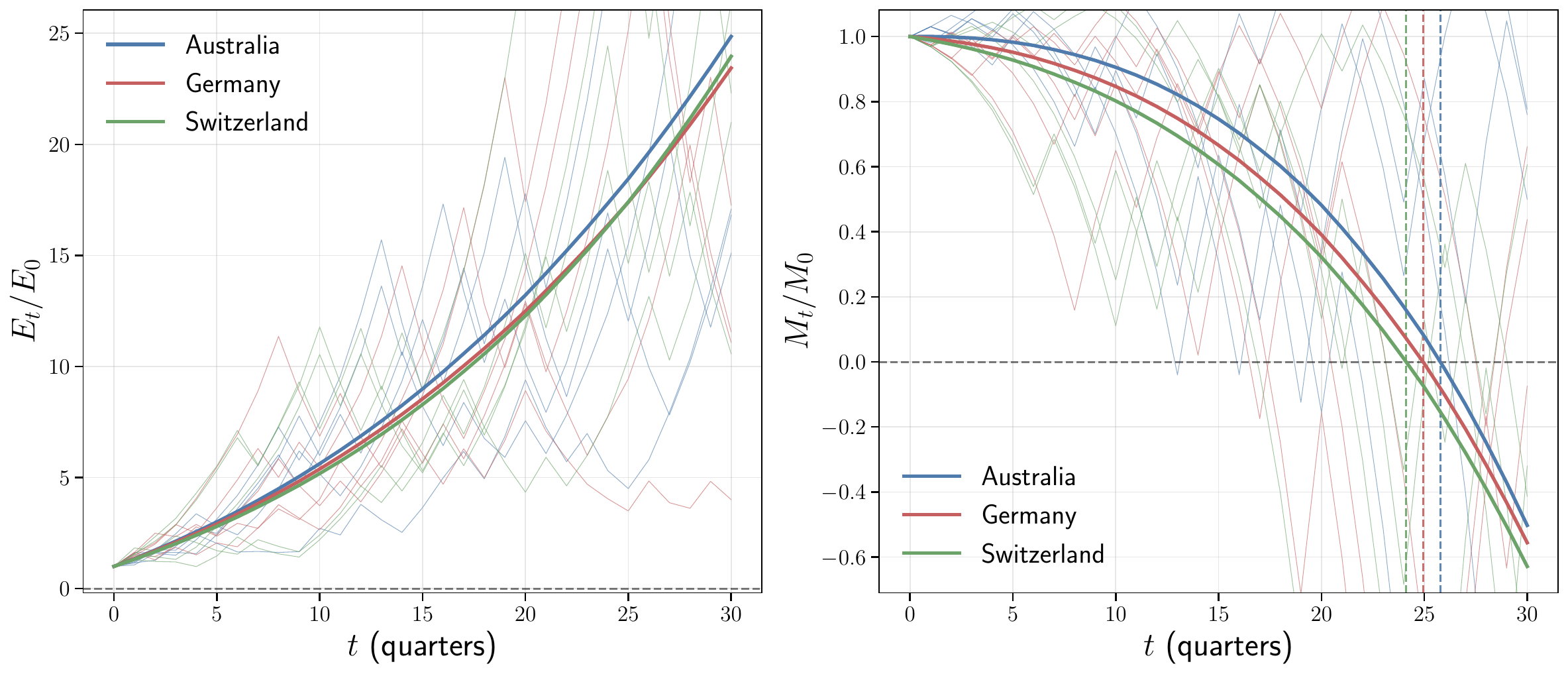}
    \caption{Calibrated Simulation: Success Scenario with  $p = 0.4$ and $s = -1.5\%$ for the Rental Case.}
    \label{fig:success_rental}
\end{figure}

Fig. \ref{fig:success_rental} shows the same country ordering as in the owner-occupied success case: Switzerland achieves mortgage repayment first, followed by Germany and Australia. Both Germany and Australia exhibit faster convergence relative to their owner-occupied scenarios.

\paragraph*{Default Scenario}

\begin{figure}
    \centering
    \includegraphics[width=\linewidth]{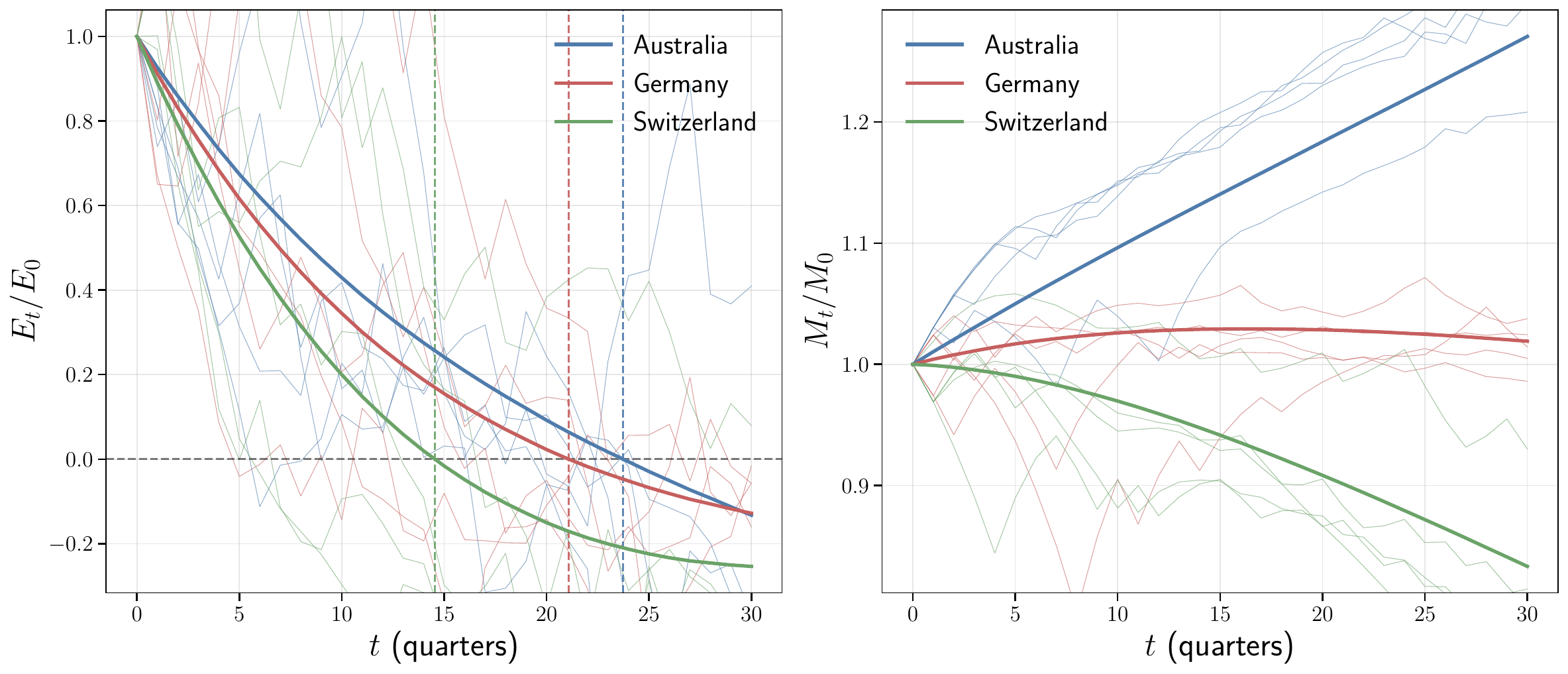}
    \caption{Calibrated Simulation: Default Scenario with  $p = 0.4$ and $s = -1.5\%$ for the Rental Case.}
    \label{fig:default rental}
\end{figure}

Fig. \ref{fig:default rental} reveals a noteworthy result: in this adverse default scenario, Switzerland reaches equity depletion first, followed by Germany and Australia, indicating a reversal of the ordering observed in owner-occupied defaults.
This counterintuitive outcome arises from the interaction between tax shields and the dominant eigenvalue of the process under adverse conditions. With $p = 0.4 (< 0.5)$, the term $\ell\mu(2p-1)$ in the $(1,1)$ entry of $\tilde{A}$ (see Eq. \ref{eq:matrix finale}) is negative, creating a contractionary force on equity. In the Swiss case, despite tax shields on both borrowing and mortgage interest, the low base rates imply that the tax benefits $\tau_m r_m M_{t-1}$ are modest. The dominant eigenvalue controlling equity dynamics under negative $s$ and $p < 0.5$ is further below unity for Switzerland than for countries with higher rates, leading to faster proportional equity decay despite lower absolute costs. In contrast, Australia's higher rates create larger interest charges in absolute terms, but the higher tax shields ($\tau_m = \tau_b = 32\%$) generate larger absolute tax benefits. The net effect in this particular adverse scenario is that Switzerland's equity, starting from the same normalised initial condition, crosses zero first compared to the other jurisdictions.

This result highlights the nonlinear interaction between interest rates, tax shields, and fluctuations in investment and housing markets. Tax shields do not uniformly improve outcomes; rather, their effectiveness depends critically on the underlying rate environment and the relative magnitudes of positive versus negative shocks. 
In favourable conditions, low rates with moderate shields dominate. In sufficiently adverse conditions, higher absolute tax benefits from higher rates can delay default longer than lower absolute benefits from lower rates.

\section{Conclusions and Policy Implications}\label{sec: concl}

This study builds upon the debt recycling model of Aufiero et al. (2025) \cite{aufiero2025phase} by incorporating mortgage interest rates, borrowing costs on equity-backed credit lines, and their associated marginal tax shields. We calibrate the extended framework to institutional parameters for Australia, Germany, and Switzerland, encompassing both owner-occupied and rental property scenarios.

Country-specific calibrations reveal systematic differences in debt recycling viability. Switzerland consistently exhibits the most favourable conditions, combining low rates with moderate tax shields, applicable to both owner-occupied and rental properties. Success regions are widest and outcome times shortest in this jurisdiction. Germany occupies an intermediate position, with moderate rates but no tax shields for owner-occupied properties, improving substantially for rental properties. Australia presents the most challenging environment for owner-occupiers, with high rates and investment-only borrowing deductibility, though the gap with other countries is narrower for rental properties.
Rental properties debt recycling outcomes consistently outperform owner-occupied housing (as shown in phase diagrams and simulations), as mortgage interest deductibility ($\tau_m > 0$) provides an equity boost unavailable to primary residence owners in Australia and Germany. This institutional asymmetry suggests that debt recycling strategies are systematically more viable for income-producing real estate than for personal residences in jurisdictions that restrict personal mortgage interest deductions.

An unexpected result emerges in the rental default scenario: Switzerland reaches equity depletion before Germany and Australia despite having the most favourable rates and shields. This counterintuitive ordering arises because the absolute magnitude of the mortgage tax benefit $\tau_m r_m M_{t-1}$ is smallest in Switzerland reflecting its low base rates. When investment returns are predominantly negative and housing depreciates, this modest benefit provides an insufficient cushion against equity erosion. Australia's higher rate environment generates a larger absolute tax benefit that delays default longer. 
This finding demonstrates that tax shields do not uniformly improve outcomes; their effectiveness depends on the interaction between rate levels, shield magnitudes, and the sign and volatility of stochastic shocks.

The permanent re-mortgaging phase, while seemingly unphysical in that it never reaches a terminal state, could actually represent a viable economic arrangement from certain perspectives: the lending institution continuously profits from sustained interest payments on both mortgage and credit line balances, while the borrower -- whose priority is wealth accumulation through leveraged investment rather than outright home ownership -- maintains access to ongoing equity extraction as long as $E_t > M_t$ holds. In this regime, the household remains a highly profitable ``evergreen'' borrower from the lender's viewpoint, servicing debt obligations without ever fully extinguishing the principal, thereby generating perpetual interest income, while maintaining adequate collateral coverage.

The results carry significant implications for household financial planning, mortgage lending practices, and regulatory oversight across jurisdictions.
Debt recycling should not be considered a universal wealth-building strategy, but rather a jurisdiction-specific and property-type-specific tactic whose viability depends critically on local interest rate levels and tax treatment. In high-rate, low-shield environments (e.g., owner-occupied properties in Australia), the strategy is viable only under exceptionally favourable investment and housing market conditions, which may not persist long enough to achieve mortgage satisfaction within realistic time horizons. Households in such environments should generally prefer traditional amortisation strategies, unless they possess strong beliefs on sustained favourable market conditions in the future. Conversely, in low-rate, high-shield environments (e.g., Switzerland, or rental properties in all three countries), debt recycling becomes viable across a broader parameter space, though success remains sensitive to market conditions and is never guaranteed.

In jurisdictions where mortgage interest is non-deductible for owner-occupiers, lenders might consider offering lower borrowing rates $r_b$ on investment credit lines to partially compensate for the absence of tax benefits and expand the feasible parameter space. Dynamic LTV caps that adjust with prevailing market conditions (tightening when $s < 0$ or when investment performance deteriorates) could prevent excessive leverage accumulation in adverse scenarios, reducing systemic default risk.

The stark differences between owner-occupied and rental outcomes within countries highlight the incentive effects of differential tax treatment. Policymakers concerned about housing affordability and household leverage might reconsider policies that favour rental property investment over owner-occupancy through asymmetric interest deductibility. Conversely, extending mortgage interest deductibility to primary residences would expand debt recycling viability for owner-occupiers, potentially accelerating wealth accumulation for homeowners, but also increasing household leverage and default risk during downturns. The Swiss model of broad interest deductibility paired with imputed rental income taxation represents one approach to equalising treatment, though it introduces administrative complexity.

Our results suggest that debt recycling creates heterogeneous risks across borrower types and jurisdictions. Regulators might develop differential capital requirements or loan-loss provisioning rules for equity-backed lending, calibrated to local interest rate and tax environments. In high-rate jurisdictions without tax shields, higher risk weights on such lending would reflect the narrower success regions and elevated default probabilities evident in our phase diagrams. Stress testing frameworks should incorporate correlated shocks to housing markets and investment returns, given that both affect equity simultaneously in debt recycling arrangements. The permanent re-mortgaging regions identified in our phase diagrams suggest potential for self-reinforcing leverage cycles that merit supervisory attention, particularly in low-rate environments where such dynamics are most likely.

The substantial international variation in outcomes demonstrates that mortgage finance best practices are not universally applicable. Regulatory frameworks and consumer protection standards developed in one jurisdiction may be inappropriate or ineffective in another with different rate and tax structures. International policy coordination efforts should account for these institutional differences rather than assuming uniform standards are optimal.

\section*{Acknowledgments} P.V. acknowledges support from UKRI Future Leaders Fellowship Scheme
(No. MR/X023028/1). F.C. acknowledges support of the Economic and Social Research Council (ESRC) in funding the Systemic Risk Centre at the LSE (ES/Y010612/1).

\end{document}